\documentclass[aps,prl,twocolumn]{revtex4}
\usepackage{graphicx}

\begin{document}

\title{Evidence for the Coexistence of Anisotropic Superconducting Gap and
Nonlocal Effects in the Non-magnetic Superconductor LuNi$_{2}$B$_{2}$C}
\author{Tuson Park$^{1}$, Elbert E. M. Chia$^{2}$, M. B. Salamon$^{2}$, E. D.
Bauer$^{1}$, I.~Vekhter$^{1*}$, J. D. Thompson$^{1}$, Eun Mi Choi$^{3}$, Heon Jung Kim$^{3}$, Sung-Ik Lee$^{3}$, and P. C. Canfield$^{4}$}
\affiliation{$^{1}$Los Alamos National Laboratory, Los Alamos, New Mexico 87545\\
$^{2}$Department of Physics and Material Research Laboratory, University of
Illinois at Urbana-Champaign, IL 61801, USA\\
$^{3}$National Creative Research Initiative Center for Superconductivity and
Department of Physics, Pohang University of Science and Technology, Pohang
790-784, Republic of Korea\\
$^{4}$Ames Laboratory, Department of Physics and Astronomy, Iowa State
University, Ames, Iowa 50011}
\date{\today}

\begin{abstract}
A study of the dependence of the heat capacity $C_{p}(\alpha)$ on field angle in LuNi$_{2}$B$_{2}$C reveals an anomalous disorder effect. For pure samples, $C_{p}(\alpha)$ exhibits a fourfold variation as the field $H < H_{c2}$ is rotated in the $[001]$ plane, with minima along $<100>$ ($\alpha = 0$). A slightly disordered sample, however, develops anomalous secondary minima along $<110>$ for $\mu_{0}H > 1$~T, leading to an 8-fold pattern at 2~K and 1.5~T. The anomalous pattern is discussed in terms of coexisting superconducting gap anisotropy and non-local effects.

\end{abstract}

\maketitle

Exotic superconductors, defined as those that follow the Uemura relation \cite{uemura91} between the superconducting transition temperature $T_{c}$ and
the magnetic penetration depth $\lambda $, $T_{c}\propto \lambda ^{-2}$, have
physical properties that differ from conventional (BCS) superconductors.
High-$T_{c}$ cuprates, bismuthates, Chevrel-phases, organic, and
heavy-fermion superconductors were suggested to constitute this class. For
most of the members, the superconducting gap function is highly anisotropic
or has gap zeros on the Fermi surface \cite{annett99}. The rare-earth nickel
borocarbides RNi$_{2}$B$_{2}$C (R=Y, Lu, Tm, Er, Ho, and Dy) remain a
challenge because they share many features in common with the exotic
superconductors but, like elemental BCS superconductors, fall below the
Uemura trend \cite{brandow03}.

Unlike BCS superconductors, which exhibit exponential temperature dependence
in density-of-states(DOS)-dependent quantities below $T_{c}$, the
borocarbides have power-law temperature dependences in their low
temperature specific heat \cite{nohara97}, NMR $1/T_{1}$ \cite{zheng98}, and
thermal conductivity \cite{boaknin01}, indicating that electronic
excitations persist even well below $T_{c}$. Recently, compelling evidence
for the presence of nodes along $<100>$ directions has been reported both
from field-angle thermal conductivity \cite{izawa02} and field-angle heat
capacity measurements \cite{tuson03} of YNi$_{2}$B$_{2}$C.

Another interesting feature of the borocarbides is that the flux-line
lattice (FLL) undergoes a field-driven transition from hexagonal to square
with increasing magnetic field $H\parallel \lbrack 001]$ \cite%
{wilde97,eskildsen97,yethiraj97}. Kogan \textit{et~al.} successfully incorporated nonlocal corrections to the London model and Fermi surface anisotropy in
these materials \cite{kogan97} to describe the transition within a BCS
scheme, i.e., isotropic s-wave gap with electron-phonon coupling. Nonlocal
effects were also used to explain, without resorting to any exotic order
parameter, the modulation of the upper critical field $H_{c2}(\alpha )$ \cite%
{metlushko97} and magnetization $M(\alpha )$ \cite{civale99} with the
angle $\alpha $ between the magnetic field and the crystal axes. The two
seemingly irreconcilable viewpoints have only increased the confusion about
the nature of the order parameter of the borocarbides. Recently, Nakai et
al. considered the coexistence of an anisotropic gap and nonlocal effects in
the borocarbides \cite{nakai02} and thereby explained the reentrant FLL
transformation \cite{eskildsen2001} in terms of the interplay between the two
effects. In this report, we present further evidence for the coexistence of
the gap anisotropy and nonlocal effects. In clean systems, the gap
anisotropy effects dominate while both effects are comparable in less clean
samples.

Single crystals of LuNi$_{2}$B$_{2}$C were grown in a Ni$_{2}$B flux as described elsewhere \cite{cho95}. Samples~\textit{A} and \textit{C} were
not annealed while sample~\textit{N} was annealed at $T=1000$ C$^{\circ }$ for 100
hours under high vacuum. Typical sample dimensions were $1\times 1\times 0.1$~mm$^{3}$. The crystal axes of the sample
were determined by two independent methods, x-ray and upper critical field
measurements as a function of magnetic field angle \cite{metlushko97}, which
were consistent with each other. The $T_{c}$'s, determined by the point
where the steepest drop occurs in the resistive superconducting transition,
are 15.5, 15.9, and 16.1~K for sample~\textit{A}, \textit{C} and \textit{N} respectively (not shown). The resistivity at $T_{c}$ is 2.34 and 1.44~$\mu \Omega \cdot $%
cm for sample~\textit{A} and sample~\textit{N}, corresponding to mean free paths of 144.5 and
234~$\mathring{A}$ respectively. Assuming that 16.1~K is the transition
temperature for a pure sample, sample~\textit{A} with $T_{c}$ of 15.5~K is
equivalent to 0.8~$\%$ of Co doping on the Ni site, i.e., Lu(Ni$_{1-x}$Co$_{x}
$)$_{2}$B$_{2}$C with $x=0.008$ \cite{cheon98}. The disorder in sample~\textit{A} may be associated with defects which can be removed by judicious post growth annealing \cite{miao02}.

Using an ac technique \cite{tuson03}, we have measured the low temperature specific heat of non-magnetic LuNi$_{2}$%
B$_{2}$C as a function of magnetic field intensity and magnetic field angle.
Three samples of LuNi$_{2}$B$_{2}$C with different $T_{c}$'s, labeled \textit{A}, \textit{C},
and \textit{N}, were studied and revealed an anomalous disorder effect. The heat
capacity of the samples with higher $T_{c}$'s shows a fourfold pattern as a
function of magnetic field angle, confirming the result from the other
non-magnetic borocarbide YNi$_{2}$B$_{2}$C that the superconducting gap is
highly anisotropic with nodes along $<100>$ \cite{tuson03}. In contrast, the
heat capacity of the disordered sample with the lowest $T_{c}$ (sample~\textit{A})
shows a dramatic change in the field-angle heat capacity above 0.8~T at 2~K:
the maxima along $<110>$ split, giving rise to minima separated by $\pi /4$
(eightfold pattern) and, further, $C_{p}$ deviates from a square-root field
dependence at the same field, labeled $H_{s1}$. The eightfold pattern and
the deviation from $H^{1/2}$ dependence in sample~\textit{A} are consistent with the coexistence of an anisotropic gap and nonlocal effects.

\begin{figure}[tbp]
\centering  \includegraphics[width=8cm,clip]{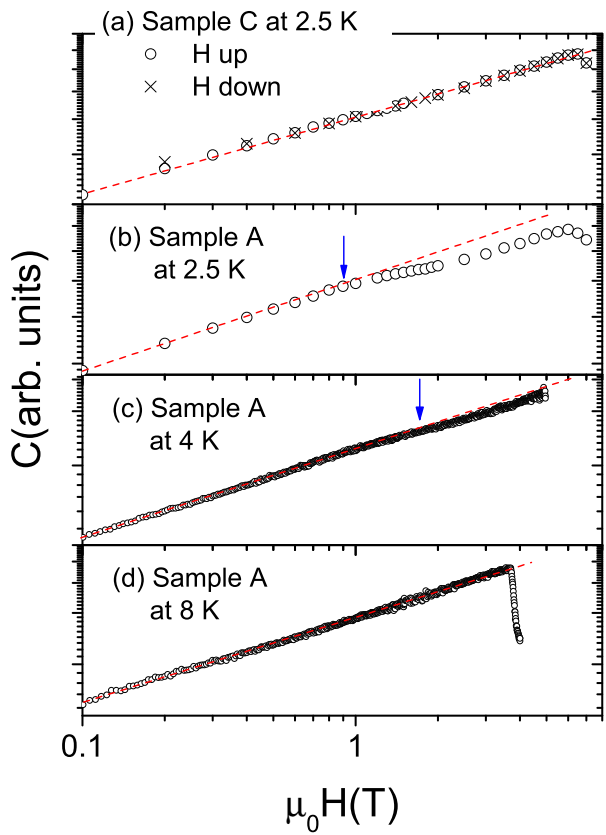}
\caption{(a) Heat capacity of sample~\textit{C} at 2.5~K for $H \parallel$[110]. (b)-(d) Heat capacity of sample~\textit{A} at 2.5, 4, and 8~K, respectively for $H \parallel$[110]. Dashed lines are $H^{1/2}$ fit and arrows indicate the deviation point from $H^{1/2}$.}
\label{figure1}
\end{figure}
Fig.~1(a) shows the magnetic field dependence of the heat
capacity of sample~\textit{C} at 2.5~K with increasing (circles) and decreasing (crosses) magnetic field. The dashed line
represents the least-square fit of $C_{0}+b(H-H_{0})^{\beta }$, where $C_{0}$
is zero-field heat capacity and $\mu_{0}H_{0}=$ 0.1~T is a fitting parameter that takes account of the Meissner effect. The best fit is obtained when $\beta
=0.46$, namely the Volovik effect for nodal superconductors \cite%
{volovik93,ichioka99}, and is consistent with previous reports on the
non-magnetic borocarbides \cite{nohara97}. Fig.~1(b), (c), and (d) shows the heat capacity of sample~\textit{A} at 2.5~K, 4~K, and 8~K, respectively. The dashed line is the square-root field dependence and the arrows indicate where the data deviate from the fit. The deviation field shows systematic increase with temperature, i.e., 0.8~T at 2.5~K, 1.8~T at 4~K, and no clear deviation at 8~K. Above the deviation field, the data fall below the $H^{1/2}$ line. 

Field-angle heat capacity directly measures the change in the DOS with magnetic field direction. The DOS of a d-wave superconductor has been shown to exhibit a fourfold oscillation with field angle against crystal axes \cite{vekhter99}. At $T=0$ K, the DOS has a
simple form: 
\begin{equation}
N(E,H,\alpha )/N_{0}\simeq D_{4}(1+\Gamma |\sin 2\alpha |),
\end{equation}%
where $D_{4}$ is a Doppler-shift induced coefficient and $\Gamma $ describes
the oscillation amplitude. The field-angle sensitive Doppler effect, arising
from the supercurrent flows circulating around the vortices, leads to maxima
in the DOS when field is along gap maxima and DOS minima when field is along
nodes. A 3D superconductor has a much reduced oscillation amplitude $%
(\approx 6\%)$ compared to that of a 2D system $(\approx 40\%)$ due to
contributions from the out-of-plane component. We note that a similar effect
is predicted for an (s+g)-wave superconductor \cite{maki03}.

Fig~2(a) and 2(b) show the low-temperature field-angle heat capacity of
sample~\textit{C} at 2.5~K and sample~\textit{A} at 2~K, respectively. The samples were
field-cooled to 2~K (or 2.5~K) and were rotated within the $ab$-plane by a
computer controlled stepping motor at increments of 3$^{\circ }$. The heat capacity with increasing and decreasing field angle showed reversible behavior for all measured fields, indicating that flux pinning effect is negligible in our measurement. Background
contributions ($C_{bkg}$) from lattice vibrations and thermometry were
subtracted in the usual manner \cite{tuson03} and the remaining
field-induced heat capacity $\Delta C=C_{total}-C_{bkg}$ was analyzed in
terms of $\Delta C(\alpha )=c(1+\Gamma |\sin 2\alpha |)$ for pure samples.
At low fields, there is a clear fourfold oscillation with minima along $<100>
$ for both samples, indicating that the zeros of the gap are located along
those directions, consistent with those of YNi$_{2}$B$_{2}$C \cite{izawa02,tuson03}. The oscillation amplitude $\Gamma $ is about 4~\% for all three samples (sample~\textit{N} is not shown, but is essentially the same as sample~\textit{C}); this is consistent with the 3D superconductivity in the borocarbides \cite{mattheis94}.
\begin{figure}[tbp]
\centering  \includegraphics[width=8.6cm,clip]{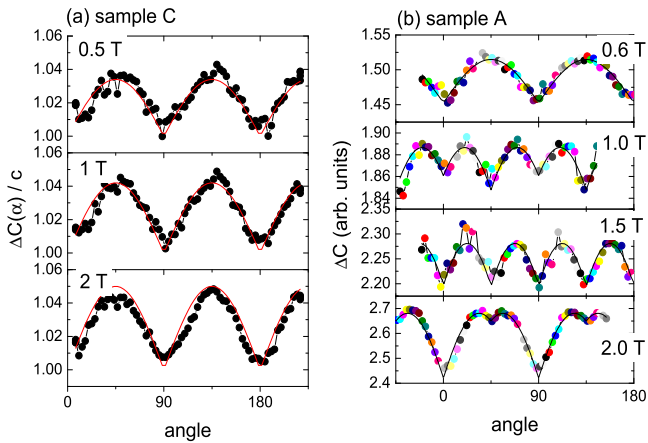}
\caption{(a) Normalized heat capacity of sample~\textit{C} at 2.5~K and a 4-fold oscillation (solid line), $\Delta C(\protect\alpha )/c=1+\Gamma |\sin 2\protect%
\alpha |$. (b) Field-induced heat capacity of sample~\textit{A} at 2~K, $%
\Delta C=C_{total}-C_{bkg}$, and a 4+4 pattern (solid line) - see text.}
\label{figure2}
\end{figure}

At 1~T, surprisingly, the heat capacity of sample~\textit{A} develops minima along $<110>$, producing two sets of fourfold patterns or 8-fold, an effect not
observed in either sample~\textit{C} or sample~\textit{N} with higher $T_{c}$ 's. The crossover field from the fourfold to the $(4+4)$ pattern of sample~\textit{A} lies between 0.6 and 1~T, which is also the point where the heat capacity of
sample~\textit{A} deviates from the square-root field dependence (see Fig.~1). With increasing field, the splitting in sample~\textit{A} gradually disappears and the field-angle heat capacity recovers its fourfold pattern above 4~T. We also
measured the field-angle heat capacity of sample~\textit{A} at 4~K to check if the
anomalous peak splitting persists at higher temperatures (not shown). The fourfold pattern now persists to 1~T, evolving into two sets of
fourfold patterns above 2~T. The 4~T data at 4~K has a shape similar to the
2~T data at 2~K. The crossover field $H_{s1}$ increases with increasing
temperature.

We focus on the fact that the anomalous 8-fold pattern occurs only at
sample~\textit{A} which has half the electronic mean free path of the sample~\textit{N} while the $T_{c}$ is slightly decreased. According to the nonlocal theory by Kogan \textit{et~al.} \cite{kogan97}, the hexagonal-to-square FLL transition depends on the electronic mean free path $l$ and the superconducting coherence length $\xi $ of the sample. Gammel \textit{et~al.} found that a mere 9\% of Co doping onto the Ni site in Lu1221 can make the FLL transition field at least 20 times higher than that of pure matrix for $H \parallel [001]$ \cite{gammel99}. Because the FLL transition field for pure sample is expected to be small \cite{kogan97}--possibly below our measurement range--the nonlocal effects would not influence the field-angle heat capacity of sample~\textit{N} (or~\textit{C}). In contrast, the disorder in sample~\textit{A} is expected to increase the transition to a higher field, i.e. to a field relevant in the field-angle heat capacity measurement.

When the magnetic field is rotated within ab-plane, the transition field may
differ with different field directions because of the different nonlocal
range, i.e., $\xi /l$. The two different transition fields can be
characterized by $H_{s1}$ and $H_{s2}$. As a magnetic field rotates within the \textit{ab}-plane for $H_{s1}\leq H\leq H_{s2}$%
, the FLL will experience a structural change (or distortion), i.e.
hexagonal for $H\parallel \lbrack 100]$ and square for $H\parallel \lbrack
110]$. Since the borocarbides have nodes on the Fermi surface, the DOS will
differ depending on the FLL structure \cite{ichioka99}. The negative deviation from the $H^{1/2}$ above $H_{s1}$ in the heat capacity of sample~\textit{A} (see Fig.~1b) also indicates that the DOS of the hexagonal FLL is larger than that of the square FLL. The additional FLL anisotropy in the DOS will modulate the gap-anisotropy oscillation and leads to a (4+4)-fold pattern.

\begin{figure}[tbp]
\centering  \includegraphics[width=8cm,clip]{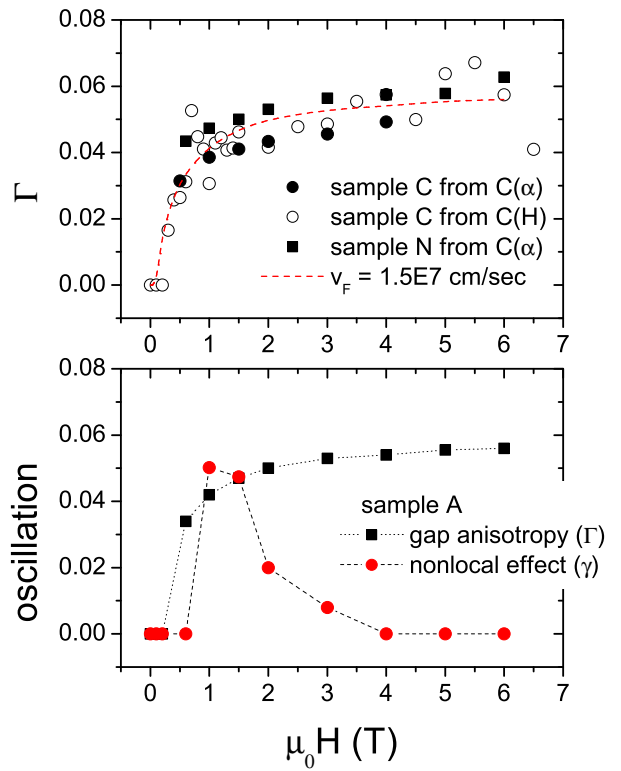}
\caption{Oscillation amplitude $\Gamma$ of sample~\textit{C} (circles) and sample~\textit{N} (squares). The dashed line is from the 3D nodal quasiparticle theory with $v_{F}=1.5\times 10^{7}$ cm/sec \cite{tuson03}. The bottom panel contrasts the oscillation amplitude
due to the nonlocal effect (circles) and gap anisotropy (squares).}
\label{figure3}
\end{figure}
We hypothesize that the two effects are independent of each other and have a
form of two cusped 4-fold oscillations in the DOS: 
\begin{equation}
\Delta C(\alpha )=p1+p2(1+\Gamma |sin2\alpha |)(1+\gamma |sin2(\alpha -45)|),
\end{equation}%
where $p1$ and $p2$ are fitting parameters. The value $\Gamma $ represents
the oscillation due to gap anisotropy in pure samples (see Fig.~3). The
nonlocal effects give rise to a 45$^{\circ }$-shifted 4-fold pattern and are
accounted for by $\gamma $. The solid line in Fig.~2(b) is the least square
fit of Eq.~(2) and represents the data very well. The oscillations due to the nonlocal effects ($\gamma $) and the gap anisotropy ($\Gamma $%
) at 2~K are compared as a function of magnetic field at the bottom panel of
Fig.~3. The FLL effect $\gamma $ increases sharply above 0.6~T and decreases
gradually to zero at 4~T, indicating that the low field corresponds to $%
H_{s1}$ and the high field to $H_{s2}$.

Fig.~4 summarizes the $H-T$ phase diagram of the disordered sample~\textit{A}. Unlike
pure samples, it has additional phase lines in the superconducting state
where the field-angle heat capacity shows a crossover from fourfold to the
(4+4)-pattern ($H_{s1}$) or vice versa ($H_{s2}$). The increase in the
crossover fields with increasing temperature is consistent with the nonlocal
effects \cite{kogan97}, adding strength to our viewpoint that the anomalous
8-fold pattern is due to the coexistence of  nonlocal effects and gap
anisotropy. We note, however, that the difference between the two transition fields $H_{s1}$ and $H_{s2}$ is larger than that expected within the nonlocal theory \cite{kogan97}. For discussion, we assume that the anisotropy in the FLL transition field between $<001>$ and $<100>$ is similar to that between $<110>$ and $<100>$ because the $H_{c2}$ anisotropy is almost same for the two configurations. The observed ratio $H_{s2} / H_{s1} \approx 4$ is larger than the predicted ratio 2 \cite{kogan97}, but is smaller than the reported ratio of 10 in YNi$_{2}$B$_{2}$C \cite{sakata00}. The difference between experiments and theory may attest that we need to account for both nonlocality and anisotropic gap nature of the borocarbides.
\begin{figure}[tbp]
\centering  \includegraphics[width=8cm,clip]{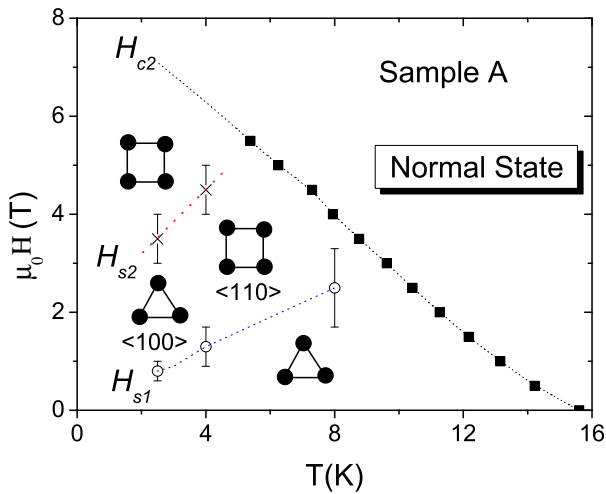}
\caption{$H-T$ phase diagram of sample~\textit{A} with lowest $T_{c}$. $H_{s1}$ is the FLL transition for $H \parallel$[110] and $H_{s1}$ for $H \parallel$[100]. The dotted lines are guide to the eye. Triangles and squares were sketched to show the corresponding FLL.}
\label{figure4}
\end{figure}

In summary, we have studied the non-magnetic superconductor LuNi$_{2}$B$_{2}$C via the dependence of heat capacity on magnetic field angle and magnetic field. Unlike pure samples, a slightly disordered sample~\textit{A} shows a deviation from $H^{1/2}$ and a (4+4)-fold pattern in $C_{p}(\alpha)$ at above 1~T. The anomalous properties were  explained in terms of the coexistence of gap anisotropy and nonlocal effects. These experiments resolve the apparently irreconcilably different views on the nature of the order parameter in the non-magnetic borocarbides.

This project was supported in part by NSF Grant No. DMR 99-72087 and at
Pohang by the Ministry of Science and Technology of Korea through the
Creative Research Initiative Program. This manuscript has been authored by Iowa State University of Science and Technology under Contract No. W-7405-ENG-82 with the U. S. Department of Energy. X-ray measurements were carried out in
the Center for Microanalysis of Materials, University of Illinois, which is
partially supported by the U.S Department of Energy under grant
DEFG02-91-ER45439. We thank S. Baily, N. Moreno-Salazar, and M. Hundley for
their help in performing experiments. T. Park and M. Salamon acknowledge benefits from the discussion with S. Balatsky, T. Leggett and K. Machida.

$^{*}$ Current address: Department of Physics and Astronomy, Louisiana State University, Baton Rouge, LA 70803, USA
\bibliography{lnbc}

\begin{thebibliography}{26}
\expandafter\ifx\csname natexlab\endcsname\relax\def\natexlab#1{#1}\fi
\expandafter\ifx\csname bibnamefont\endcsname\relax
  \def\bibnamefont#1{#1}\fi
\expandafter\ifx\csname bibfnamefont\endcsname\relax
  \def\bibfnamefont#1{#1}\fi
\expandafter\ifx\csname citenamefont\endcsname\relax
  \def\citenamefont#1{#1}\fi
\expandafter\ifx\csname url\endcsname\relax
  \def\url#1{\texttt{#1}}\fi
\expandafter\ifx\csname urlprefix\endcsname\relax\def\urlprefix{URL }\fi
\providecommand{\bibinfo}[2]{#2}
\providecommand{\eprint}[2][]{\url{#2}}

\bibitem[{\citenamefont{Uemura et~al.}(1991)\citenamefont{Uemura, Le, Luke,
  Sternlieb, Wu, Brewer, Riseman, Seaman, Maple, Ishikawa et~al.}}]{uemura91}
\bibinfo{author}{\bibfnamefont{Y.~J.} \bibnamefont{Uemura}},
  \bibinfo{author}{\bibfnamefont{L.~P.} \bibnamefont{Le}},
  \bibinfo{author}{\bibfnamefont{G.~M.} \bibnamefont{Luke}},
  \bibinfo{author}{\bibfnamefont{B.~J.} \bibnamefont{Sternlieb}},
  \bibinfo{author}{\bibfnamefont{W.~D.} \bibnamefont{Wu}},
  \bibinfo{author}{\bibfnamefont{J.~H.} \bibnamefont{Brewer}},
  \bibinfo{author}{\bibfnamefont{T.~M.} \bibnamefont{Riseman}},
  \bibinfo{author}{\bibfnamefont{C.~L.} \bibnamefont{Seaman}},
  \bibinfo{author}{\bibfnamefont{M.~B.} \bibnamefont{Maple}},
  \bibinfo{author}{\bibfnamefont{M.}~\bibnamefont{Ishikawa}},
  \bibnamefont{et~al.}, \bibinfo{journal}{Phys. Rev. Lett.}
  \textbf{\bibinfo{volume}{66}}, \bibinfo{pages}{2665} (\bibinfo{year}{1991}).

\bibitem[{\citenamefont{Annett}(1999)}]{annett99}
\bibinfo{author}{\bibfnamefont{J.~F.} \bibnamefont{Annett}},
  \bibinfo{journal}{Physica C} \textbf{\bibinfo{volume}{317}},
  \bibinfo{pages}{1} (\bibinfo{year}{1999}).

\bibitem[{\citenamefont{Brandow}(2003)}]{brandow03}
\bibinfo{author}{\bibfnamefont{B.~H.} \bibnamefont{Brandow}},
  \bibinfo{journal}{Phil. Mag.} \textbf{\bibinfo{volume}{83}},
  \bibinfo{pages}{2487} (\bibinfo{year}{2003}), \bibinfo{note}{references
  therein}.

\bibitem[{\citenamefont{Nohara et~al.}(1997)\citenamefont{Nohara, Isshiki,
  Takagi, and Cava}}]{nohara97}
\bibinfo{author}{\bibfnamefont{M.}~\bibnamefont{Nohara}},
  \bibinfo{author}{\bibfnamefont{M.}~\bibnamefont{Isshiki}},
  \bibinfo{author}{\bibfnamefont{H.}~\bibnamefont{Takagi}}, \bibnamefont{and}
  \bibinfo{author}{\bibfnamefont{R.~J.} \bibnamefont{Cava}},
  \bibinfo{journal}{J. Phys. Soc. Jpn.} \textbf{\bibinfo{volume}{66}},
  \bibinfo{pages}{1888} (\bibinfo{year}{1997}).

\bibitem[{\citenamefont{Zheng et~al.}(1998)\citenamefont{Zheng, Wada,
  Hashimoto, Kitaoka, Asayama, Takeya, and Kadowaki}}]{zheng98}
\bibinfo{author}{\bibfnamefont{G.-Q.} \bibnamefont{Zheng}},
  \bibinfo{author}{\bibfnamefont{Y.}~\bibnamefont{Wada}},
  \bibinfo{author}{\bibfnamefont{K.}~\bibnamefont{Hashimoto}},
  \bibinfo{author}{\bibfnamefont{Y.}~\bibnamefont{Kitaoka}},
  \bibinfo{author}{\bibfnamefont{K.}~\bibnamefont{Asayama}},
  \bibinfo{author}{\bibfnamefont{H.}~\bibnamefont{Takeya}}, \bibnamefont{and}
  \bibinfo{author}{\bibfnamefont{K.}~\bibnamefont{Kadowaki}},
  \bibinfo{journal}{J. Phys. Chem. Solids} \textbf{\bibinfo{volume}{59}},
  \bibinfo{pages}{2169} (\bibinfo{year}{1998}).

\bibitem[{\citenamefont{Boaknin et~al.}(2001)\citenamefont{Boaknin, Hill,
  Proust, Lupien, Taillefer, and Canfield}}]{boaknin01}
\bibinfo{author}{\bibfnamefont{E.}~\bibnamefont{Boaknin}},
  \bibinfo{author}{\bibfnamefont{R.~W.} \bibnamefont{Hill}},
  \bibinfo{author}{\bibfnamefont{C.}~\bibnamefont{Proust}},
  \bibinfo{author}{\bibfnamefont{C.}~\bibnamefont{Lupien}},
  \bibinfo{author}{\bibfnamefont{L.}~\bibnamefont{Taillefer}},
  \bibnamefont{and} \bibinfo{author}{\bibfnamefont{P.~C.}
  \bibnamefont{Canfield}}, \bibinfo{journal}{Phys. Rev. Lett.}
  \textbf{\bibinfo{volume}{87}}, \bibinfo{pages}{237001}
  (\bibinfo{year}{2001}).

\bibitem[{\citenamefont{Izawa et~al.}(2002)\citenamefont{Izawa, Kamata,
  Nakajima, Matsuda, Watanabe, Nohara, Takagi, Thalmeier, and Maki}}]{izawa02}
\bibinfo{author}{\bibfnamefont{K.}~\bibnamefont{Izawa}},
  \bibinfo{author}{\bibfnamefont{K.}~\bibnamefont{Kamata}},
  \bibinfo{author}{\bibfnamefont{Y.}~\bibnamefont{Nakajima}},
  \bibinfo{author}{\bibfnamefont{Y.}~\bibnamefont{Matsuda}},
  \bibinfo{author}{\bibfnamefont{T.}~\bibnamefont{Watanabe}},
  \bibinfo{author}{\bibfnamefont{M.}~\bibnamefont{Nohara}},
  \bibinfo{author}{\bibfnamefont{H.}~\bibnamefont{Takagi}},
  \bibinfo{author}{\bibfnamefont{P.}~\bibnamefont{Thalmeier}},
  \bibnamefont{and} \bibinfo{author}{\bibfnamefont{K.}~\bibnamefont{Maki}},
  \bibinfo{journal}{Phys. Rev. Lett.} \textbf{\bibinfo{volume}{89}},
  \bibinfo{pages}{137006} (\bibinfo{year}{2002}).

\bibitem[{\citenamefont{Park et~al.}(2003)\citenamefont{Park, Salamon, Choi,
  Kim, and Lee}}]{tuson03}
\bibinfo{author}{\bibfnamefont{T.}~\bibnamefont{Park}},
  \bibinfo{author}{\bibfnamefont{M.~B.} \bibnamefont{Salamon}},
  \bibinfo{author}{\bibfnamefont{E.~M.} \bibnamefont{Choi}},
  \bibinfo{author}{\bibfnamefont{H.~J.} \bibnamefont{Kim}}, \bibnamefont{and}
  \bibinfo{author}{\bibfnamefont{S.-I.} \bibnamefont{Lee}},
  \bibinfo{journal}{Phys. Rev. Lett.} \textbf{\bibinfo{volume}{90}},
  \bibinfo{pages}{177001} (\bibinfo{year}{2003}).

\bibitem[{\citenamefont{DeWilde et~al.}(1997)\citenamefont{DeWilde, Iavarone,
  Welp, Metlushko, Koshelev, Aranson, Crabtree, and Canfield}}]{wilde97}
\bibinfo{author}{\bibfnamefont{Y.}~\bibnamefont{DeWilde}},
  \bibinfo{author}{\bibfnamefont{M.}~\bibnamefont{Iavarone}},
  \bibinfo{author}{\bibfnamefont{U.}~\bibnamefont{Welp}},
  \bibinfo{author}{\bibfnamefont{V.}~\bibnamefont{Metlushko}},
  \bibinfo{author}{\bibfnamefont{A.~E.} \bibnamefont{Koshelev}},
  \bibinfo{author}{\bibfnamefont{I.}~\bibnamefont{Aranson}},
  \bibinfo{author}{\bibfnamefont{G.~W.} \bibnamefont{Crabtree}},
  \bibnamefont{and} \bibinfo{author}{\bibfnamefont{P.~C.}
  \bibnamefont{Canfield}}, \bibinfo{journal}{Phys. Rev. Lett.}
  \textbf{\bibinfo{volume}{78}}, \bibinfo{pages}{4273} (\bibinfo{year}{1997}).

\bibitem[{\citenamefont{Eskildsen et~al.}(1997)\citenamefont{Eskildsen, Gammel,
  Barber, Ramirez, Bishop, Andersen, Mortensen, Bolle, Lieber, and
  Canfield}}]{eskildsen97}
\bibinfo{author}{\bibfnamefont{M.~R.} \bibnamefont{Eskildsen}},
  \bibinfo{author}{\bibfnamefont{P.~L.} \bibnamefont{Gammel}},
  \bibinfo{author}{\bibfnamefont{B.~P.} \bibnamefont{Barber}},
  \bibinfo{author}{\bibfnamefont{A.~P.} \bibnamefont{Ramirez}},
  \bibinfo{author}{\bibfnamefont{D.~J.} \bibnamefont{Bishop}},
  \bibinfo{author}{\bibfnamefont{N.~H.} \bibnamefont{Andersen}},
  \bibinfo{author}{\bibfnamefont{K.}~\bibnamefont{Mortensen}},
  \bibinfo{author}{\bibfnamefont{C.~A.} \bibnamefont{Bolle}},
  \bibinfo{author}{\bibfnamefont{C.~M.} \bibnamefont{Lieber}},
  \bibnamefont{and} \bibinfo{author}{\bibfnamefont{P.~C.}
  \bibnamefont{Canfield}}, \bibinfo{journal}{Phys. Rev. Lett.}
  \textbf{\bibinfo{volume}{79}}, \bibinfo{pages}{487} (\bibinfo{year}{1997}).

\bibitem[{\citenamefont{Yethiraj et~al.}(1997)\citenamefont{Yethiraj, Paul,
  Tomy, and Forgan}}]{yethiraj97}
\bibinfo{author}{\bibfnamefont{M.}~\bibnamefont{Yethiraj}},
  \bibinfo{author}{\bibfnamefont{D.~M.} \bibnamefont{Paul}},
  \bibinfo{author}{\bibfnamefont{C.~V.} \bibnamefont{Tomy}}, \bibnamefont{and}
  \bibinfo{author}{\bibfnamefont{E.~M.} \bibnamefont{Forgan}},
  \bibinfo{journal}{Phys. Rev. Lett.} \textbf{\bibinfo{volume}{78}},
  \bibinfo{pages}{4849} (\bibinfo{year}{1997}).

\bibitem[{\citenamefont{Kogan et~al.}(1997)\citenamefont{Kogan, Bullock,
  Harmon, Miranovic, Dobrosavljevic-Grujic, Gammel, and Bishop}}]{kogan97}
\bibinfo{author}{\bibfnamefont{V.~G.} \bibnamefont{Kogan}},
  \bibinfo{author}{\bibfnamefont{M.}~\bibnamefont{Bullock}},
  \bibinfo{author}{\bibfnamefont{B.}~\bibnamefont{Harmon}},
  \bibinfo{author}{\bibfnamefont{P.}~\bibnamefont{Miranovic}},
  \bibinfo{author}{\bibfnamefont{L.}~\bibnamefont{Dobrosavljevic-Grujic}},
  \bibinfo{author}{\bibfnamefont{P.~L.} \bibnamefont{Gammel}},
  \bibnamefont{and} \bibinfo{author}{\bibfnamefont{D.~J.}
  \bibnamefont{Bishop}}, \bibinfo{journal}{Phys. Rev. B}
  \textbf{\bibinfo{volume}{55}}, \bibinfo{pages}{R8693} (\bibinfo{year}{1997}).

\bibitem[{\citenamefont{Metlushko et~al.}(1997)\citenamefont{Metlushko, Welp,
  Koshelev, Aranson, Crabtree, and Canfield}}]{metlushko97}
\bibinfo{author}{\bibfnamefont{V.}~\bibnamefont{Metlushko}},
  \bibinfo{author}{\bibfnamefont{U.}~\bibnamefont{Welp}},
  \bibinfo{author}{\bibfnamefont{A.}~\bibnamefont{Koshelev}},
  \bibinfo{author}{\bibfnamefont{I.}~\bibnamefont{Aranson}},
  \bibinfo{author}{\bibfnamefont{G.~W.} \bibnamefont{Crabtree}},
  \bibnamefont{and} \bibinfo{author}{\bibfnamefont{P.~C.}
  \bibnamefont{Canfield}}, \bibinfo{journal}{Phys. Rev. Lett.}
  \textbf{\bibinfo{volume}{79}}, \bibinfo{pages}{1738} (\bibinfo{year}{1997}).

\bibitem[{\citenamefont{Civale et~al.}(1999)\citenamefont{Civale, Silhanek,
  Thompson, Song, Tomy, and Paul}}]{civale99}
\bibinfo{author}{\bibfnamefont{L.}~\bibnamefont{Civale}},
  \bibinfo{author}{\bibfnamefont{A.~V.} \bibnamefont{Silhanek}},
  \bibinfo{author}{\bibfnamefont{J.~R.} \bibnamefont{Thompson}},
  \bibinfo{author}{\bibfnamefont{K.~J.} \bibnamefont{Song}},
  \bibinfo{author}{\bibfnamefont{C.~V.} \bibnamefont{Tomy}}, \bibnamefont{and}
  \bibinfo{author}{\bibfnamefont{D.~M.} \bibnamefont{Paul}},
  \bibinfo{journal}{Phys. Rev. Lett.} \textbf{\bibinfo{volume}{83}},
  \bibinfo{pages}{3920} (\bibinfo{year}{1999}).

\bibitem[{\citenamefont{Nakai et~al.}(2002)\citenamefont{Nakai, Miranovic,
  Ichioka, and Machida}}]{nakai02}
\bibinfo{author}{\bibfnamefont{N.}~\bibnamefont{Nakai}},
  \bibinfo{author}{\bibfnamefont{P.}~\bibnamefont{Miranovic}},
  \bibinfo{author}{\bibfnamefont{M.}~\bibnamefont{Ichioka}}, \bibnamefont{and}
  \bibinfo{author}{\bibfnamefont{K.}~\bibnamefont{Machida}},
  \bibinfo{journal}{Phys. Rev. Lett} \textbf{\bibinfo{volume}{89}},
  \bibinfo{pages}{237004} (\bibinfo{year}{2002}).

\bibitem[{\citenamefont{Eskildsen et~al.}(2001)\citenamefont{Eskildsen,
  Abrahamsen, Kogan, Gammel, Mortensen, Andersen, and
  Canfield}}]{eskildsen2001}
\bibinfo{author}{\bibfnamefont{M.~R.} \bibnamefont{Eskildsen}},
  \bibinfo{author}{\bibfnamefont{A.~B.} \bibnamefont{Abrahamsen}},
  \bibinfo{author}{\bibfnamefont{V.~G.} \bibnamefont{Kogan}},
  \bibinfo{author}{\bibfnamefont{P.~L.} \bibnamefont{Gammel}},
  \bibinfo{author}{\bibfnamefont{K.}~\bibnamefont{Mortensen}},
  \bibinfo{author}{\bibfnamefont{N.~H.} \bibnamefont{Andersen}},
  \bibnamefont{and} \bibinfo{author}{\bibfnamefont{P.~C.}
  \bibnamefont{Canfield}}, \bibinfo{journal}{Phys. Rev. Lett.}
  \textbf{\bibinfo{volume}{86}}, \bibinfo{pages}{5148} (\bibinfo{year}{2001}).

\bibitem[{\citenamefont{Cho et~al.}(1995)\citenamefont{Cho, Canfield, Miller,
  Johnston, Beyermann, and Yatskar}}]{cho95}
\bibinfo{author}{\bibfnamefont{B.~K.} \bibnamefont{Cho}},
  \bibinfo{author}{\bibfnamefont{P.~C.} \bibnamefont{Canfield}},
  \bibinfo{author}{\bibfnamefont{L.~L.} \bibnamefont{Miller}},
  \bibinfo{author}{\bibfnamefont{D.~C.} \bibnamefont{Johnston}},
  \bibinfo{author}{\bibfnamefont{W.~P.} \bibnamefont{Beyermann}},
  \bibnamefont{and} \bibinfo{author}{\bibfnamefont{A.}~\bibnamefont{Yatskar}},
  \bibinfo{journal}{Phys. Rev. B} \textbf{\bibinfo{volume}{52}},
  \bibinfo{pages}{3684} (\bibinfo{year}{1995}).

\bibitem[{\citenamefont{Cheon et~al.}(1998)\citenamefont{Cheon, Fisher, Kogan,
  Canfield, Miranovic, and Gammel}}]{cheon98}
\bibinfo{author}{\bibfnamefont{K.~O.} \bibnamefont{Cheon}},
  \bibinfo{author}{\bibfnamefont{I.~R.} \bibnamefont{Fisher}},
  \bibinfo{author}{\bibfnamefont{V.~G.} \bibnamefont{Kogan}},
  \bibinfo{author}{\bibfnamefont{P.~C.} \bibnamefont{Canfield}},
  \bibinfo{author}{\bibfnamefont{P.}~\bibnamefont{Miranovic}},
  \bibnamefont{and} \bibinfo{author}{\bibfnamefont{P.~L.}
  \bibnamefont{Gammel}}, \bibinfo{journal}{Phys. Rev. B}
  \textbf{\bibinfo{volume}{58}}, \bibinfo{pages}{6463} (\bibinfo{year}{1998}).

\bibitem[{\citenamefont{Miao et~al.}(2002)\citenamefont{Miao, Bud'ko, and
  Canfield}}]{miao02}
\bibinfo{author}{\bibfnamefont{X.~Y.} \bibnamefont{Miao}},
  \bibinfo{author}{\bibfnamefont{S.~L.} \bibnamefont{Bud'ko}},
  \bibnamefont{and} \bibinfo{author}{\bibfnamefont{P.~C.}
  \bibnamefont{Canfield}}, \bibinfo{journal}{J. Alloys Comp.}
  \textbf{\bibinfo{volume}{338}}, \bibinfo{pages}{13} (\bibinfo{year}{2002}).

\bibitem[{\citenamefont{Volovik}(1993)}]{volovik93}
\bibinfo{author}{\bibfnamefont{G.~E.} \bibnamefont{Volovik}},
  \bibinfo{journal}{JETP Lett.} \textbf{\bibinfo{volume}{58}},
  \bibinfo{pages}{469} (\bibinfo{year}{1993}).

\bibitem[{\citenamefont{Ichioka et~al.}(1999)\citenamefont{Ichioka, Hasegawa,
  and Machida}}]{ichioka99}
\bibinfo{author}{\bibfnamefont{M.}~\bibnamefont{Ichioka}},
  \bibinfo{author}{\bibfnamefont{A.}~\bibnamefont{Hasegawa}}, \bibnamefont{and}
  \bibinfo{author}{\bibfnamefont{K.}~\bibnamefont{Machida}},
  \bibinfo{journal}{Phys. Rev. B} \textbf{\bibinfo{volume}{59}},
  \bibinfo{pages}{184} (\bibinfo{year}{1999}).

\bibitem[{\citenamefont{Vekhter et~al.}(1999)\citenamefont{Vekhter, Hirschfeld,
  Carbotte, and Nicol}}]{vekhter99}
\bibinfo{author}{\bibfnamefont{I.}~\bibnamefont{Vekhter}},
  \bibinfo{author}{\bibfnamefont{P.~J.} \bibnamefont{Hirschfeld}},
  \bibinfo{author}{\bibfnamefont{J.~P.} \bibnamefont{Carbotte}},
  \bibnamefont{and} \bibinfo{author}{\bibfnamefont{E.~J.} \bibnamefont{Nicol}},
  \bibinfo{journal}{Phys. Rev. B} \textbf{\bibinfo{volume}{59}},
  \bibinfo{pages}{R9023} (\bibinfo{year}{1999}).

\bibitem[{\citenamefont{Maki et~al.}(2003)\citenamefont{Maki, Won, Thalmeier,
  Yuan, Izawa, and Matsuda}}]{maki03}
\bibinfo{author}{\bibfnamefont{K.}~\bibnamefont{Maki}},
  \bibinfo{author}{\bibfnamefont{H.}~\bibnamefont{Won}},
  \bibinfo{author}{\bibfnamefont{P.}~\bibnamefont{Thalmeier}},
  \bibinfo{author}{\bibfnamefont{Q.}~\bibnamefont{Yuan}},
  \bibinfo{author}{\bibfnamefont{K.}~\bibnamefont{Izawa}}, \bibnamefont{and}
  \bibinfo{author}{\bibfnamefont{Y.}~\bibnamefont{Matsuda}},
  \bibinfo{journal}{Europhys. Lett.} \textbf{\bibinfo{volume}{64}},
  \bibinfo{pages}{496} (\bibinfo{year}{2003}).

\bibitem[{\citenamefont{Mattheiss}(1994)}]{mattheis94}
\bibinfo{author}{\bibfnamefont{L.~F.} \bibnamefont{Mattheiss}},
  \bibinfo{journal}{Phys. Rev. B} \textbf{\bibinfo{volume}{49}},
  \bibinfo{pages}{13279} (\bibinfo{year}{1994}).

\bibitem[{\citenamefont{Gammel et~al.}(1999)\citenamefont{Gammel, Bishop,
  Eskildsen, Mortensen, Andsersen, Fisher, Cheon, Canfield, and
  Kogan}}]{gammel99}
\bibinfo{author}{\bibfnamefont{P.~L.} \bibnamefont{Gammel}},
  \bibinfo{author}{\bibfnamefont{D.~J.} \bibnamefont{Bishop}},
  \bibinfo{author}{\bibfnamefont{M.~R.} \bibnamefont{Eskildsen}},
  \bibinfo{author}{\bibfnamefont{K.}~\bibnamefont{Mortensen}},
  \bibinfo{author}{\bibfnamefont{N.~H.} \bibnamefont{Andsersen}},
  \bibinfo{author}{\bibfnamefont{I.~R.} \bibnamefont{Fisher}},
  \bibinfo{author}{\bibfnamefont{K.~O.} \bibnamefont{Cheon}},
  \bibinfo{author}{\bibfnamefont{P.~C.} \bibnamefont{Canfield}},
  \bibnamefont{and} \bibinfo{author}{\bibfnamefont{V.~G.} \bibnamefont{Kogan}},
  \bibinfo{journal}{Phys. Rev. Lett.} \textbf{\bibinfo{volume}{82}},
  \bibinfo{pages}{4082} (\bibinfo{year}{1999}).

\bibitem[{\citenamefont{Sakata et~al.}(2000)\citenamefont{Sakata, Oosawa,
  Matsuba, Nishida, Takeya, and Hirata}}]{sakata00}
\bibinfo{author}{\bibfnamefont{H.}~\bibnamefont{Sakata}},
  \bibinfo{author}{\bibfnamefont{M.}~\bibnamefont{Oosawa}},
  \bibinfo{author}{\bibfnamefont{K.}~\bibnamefont{Matsuba}},
  \bibinfo{author}{\bibfnamefont{N.}~\bibnamefont{Nishida}},
  \bibinfo{author}{\bibfnamefont{H.}~\bibnamefont{Takeya}}, \bibnamefont{and}
  \bibinfo{author}{\bibfnamefont{K.}~\bibnamefont{Hirata}},
  \bibinfo{journal}{Phys. Rev. Lett.} \textbf{\bibinfo{volume}{84}},
  \bibinfo{pages}{1583} (\bibinfo{year}{2000}).

\end{thebibliography}

\end{document}